\documentclass[amssymb,amsmath,aps,showpacs,twocolumn,prd]{revtex4}
\usepackage{graphicx}
\def\beq{\begin{equation}}
\def\eeq{\end{equation}}
\def\be{\begin{equation}}
\def\ee{\end{equation}}
\def\bea{\begin{eqnarray}}
\def\eea{\end{eqnarray}}
\def\nnb{\nonumber}

\newcommand{\gsim}{\lower.7ex\hbox{$\;\stackrel{\textstyle>}{\sim}\;$}}
\newcommand{\lsim}{\lower.7ex\hbox{$\;\stackrel{\textstyle<}{\sim}\;$}}

\begin{document}

 \title{ Generation and search of axion-like light particle using intense
  crystalline field }
 \author{ Wei Liao}
 \affiliation{
  Center for High Energy Physics, Peking University,
 Beijing 100871, P. R. China \\
 \vskip 0.1cm
  Institute of Modern Physics,
 East China University of Science and Technology, \\
 130 Meilong Road, Shanghai 200237, P.R. China 
}


\begin{abstract}
 Intense electric field $\sim 10^{10}-10^{11}$ V$/$cm in crystal has been 
 known for a long time and has wide applications. We study the conversion 
 of axion-like light particle and photon in the intense electric field in 
 crystal. We find that the conversion of axion-like particle and photon 
 happens for energy larger than keV range. We propose search of axion-like 
 light particle using the intense crystalline field. We discuss the solar 
 axion search experiment and a variety of shining-through-wall experiment using 
 crystalline field. Due to the intense crystalline field which corresponds 
 to magnetic field $\sim 10^4-10^5$ Tesla these experiments are very
 interesting. In particular these experiments can probe the mass range of 
 axion-like particle from eV to keV.
\end{abstract}
 \maketitle

 The search of
 axion-like particle(ALP) has attracted a lot of interests 
 after the axion was invented in a mechanism solving the strong CP
 problem in the Standard Model~\cite{axion1,axion2, axion3,axion4}. 
 The search of ALP is based on a mechanism  that
 ALP and photon can transform to each other in 
 external electromagnetic field, e.g. in external magnetic 
 field~\cite{sikivie}. Many varieties of ALP search experiments 
 based on this mechanism have been proposed and been done
 ~\cite{review, review1}. So far, no evidence of ALP
 has been found. One of the crucial difficulties in laboratory search of ALP
 is that the external magnetic field is 
 maximally around a few Tesla in laboratory.

 It is well known for a long time that extremely strong electric field
 exists in crystal.
 For a charged particle incident on a crystal with a direction
 almost parallel to a crystallographic plane the strong
 electric field of nuclei add constructively so that a macroscopic
 and continuous electric field, the plane continuum electric field, is obtained. 
 As a consequence, positrons can be trapped in a potential well of 
 depth $\sim 10^2$ V between two crystallographic planes with 
 lattice distance $\sim 0.1$ nm and are channeled 
 between two parallel crystallographic planes. 
 The plane continuum electric field 
 is estimated $\sim 10^{10}-10^{11}$ V$/$cm which corresponds to 
 magnetic field $\sim 10^{4}-10^5$ T. 

 Charged particles when channeling in crystal oscillate in the transverse
 direction and
 can produce coherent radiation which has been intensively explored 
 theoretically and experimentally~\cite{BKSbook,Uggerhoj}.
 A high energy photon ($\gsim $ GeV) incident on a crystal is also expected to 
 be affected by the strong crystalline field and to produce electron-positron 
 pairs~\cite{Uggerhoj} since 
 the radiation and the pair-production processes
 are related by the crossing symmetry.
 It's natural to expect other quantum
 process of photon, such as the Primakoff type effect of photon-ALP
 conversion, to happen in the crystalline field. 

 In this Letter we study the ALP-photon conversion 
 in intense crystalline field and propose to do
 laboratory search of ALP using the crystalline field.
 We first briefly review the
 intense electric field in crystal.
 We study the ALP-photon conversion in crystalline field.
 We propose to do ALP search experiment
 using intense crystalline field.
 
\begin{figure}[tb]
\begin{center}
\begin{tabular}{cc}
\includegraphics[scale=1.1,width=7cm]{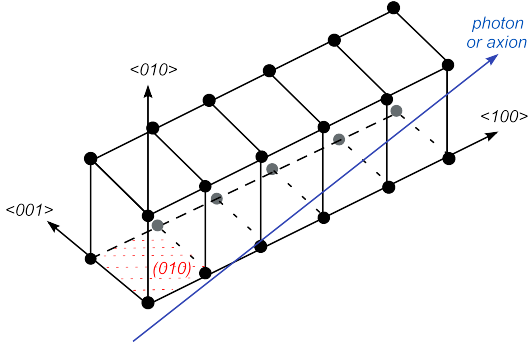}
\end{tabular}
\end{center}
\caption{Crystal lattice} \label{crystal}
\end{figure}

 The structure of a single crystal is described by the Bravais
 lattice
 \bea
 {\vec R}=l_1 {\vec a}_1+l_2 {\vec a}_2+l_3 {\vec a}_3, \label{Bravais}
 \eea
 where ${\vec a}_i$ is a primitive vector and $l_i$ an integer.
 The crystal is periodic, e.g. the atomic density $\rho({\vec x}+{\vec R})
 =\rho({\vec x})$. The same is true for the electric field and the
 potential in crystal. Using the reciprocal lattice vector ${\vec q}$
 the potential and the electric field can be Fourier transformed 
 and be expressed as
 \bea
 U({\vec x}) = \sum_{{\vec q}} U_{{\vec q}} 
 ~e^{-i{\vec q}\cdot {\vec x}},
 \label{potential1} \\
 {\vec E}({\vec x}) = \sum_{{\vec q}} {\vec E}_{{\vec q}} 
  ~e^{-i{\vec q}\cdot {\vec x}},
 \label{e-field}
 \eea
 where ${\vec E}_{\vec q}=-i {\vec q} ~U_{\vec q}$.
 ${\vec q}=2 \pi \sum_{i=1}^3 n_i {\vec b}_i$ where $n_i$ is an integer,
 ${\vec b}_i=\frac{1}{2}\epsilon_{ijk}({\vec a}_j \times {\vec a}_k)
 /({\vec a}_1\cdot({\vec a}_2\times {\vec a}_3))$ and 
 ${\vec a}_i \cdot {\vec b}_j  =\delta_{ij}$.
 In Fig. \ref{crystal} we give a plot for the cubic lattice. 
 A crystallographic plane labeled $(010)$ is shown in the
 plot. The plane is parallel to the $\langle 100 \rangle$ and 
 $\langle 001 \rangle$ axis and perpendicular to $\langle 010 \rangle$ axis.

 In a good approximation
 the plane continuum electric field can be considered constant in the longitudinal
 direction and periodic in the transverse direction. For example,
 the plane continuum potential can be expressed as~\cite{BKSbook}
 \bea
 U(y)=V[cosh(\delta(\sqrt{1+\eta^2}-\sqrt{s+\eta^2}))-1],
 \label{potential2}
 \eea
 where $s=2 |y'|/d$ and $|s| \le 1$. $d$ is the interplanar distance 
 and $y'=y-y_{mid}$ the transverse coordinate
 relative to the middle between two neighboring planes $y_{mid}$.
 $\eta$, $V$, $\delta$ are parameters.
 For example, $\delta=3.85$, $V=6.4$ V, $\eta^2=0.0007$, $d=0.119$ nm 
 for the $(110)$ plane of the W crystal at room temperature~\cite{BKSbook}. 
 The potential height $U_0$ is around $130$ V.
 For $(110)$ plane of the Ge crystal,
 $\delta=3.2$, $V=4.5$ V, $\eta^2=0.0052$, $d=0.2$ nm, $U_0=40 $ V.
 The plane continuum electric field can be derived from (\ref{potential2}).

 The Lagrangian of the photon and the ALP is
 \bea
 {\cal L} &&=\frac{1}{2}\partial_\mu \phi \partial^\mu \phi
 -\frac{1}{2} m^2_\phi \phi^2 \nnb \\
 && -\frac{1}{4} F^{\mu \nu} F_{\mu \nu} 
 +\frac{1}{4}g_\phi \phi F_{\mu \nu} F^{\mu \nu} , \label{Lag}
 \eea
 where $\phi$ is the field of ALP, $m_\phi$ the mass of ALP,
 $F^{\mu \nu}$ the field strength of photon and 
 $F^{\mu \nu} =\frac{1}{2} \varepsilon^{\mu \nu \rho \sigma} F^{\rho \sigma}$.
 $g_\phi$ is the coupling constant.

 For a uniform incident flux 
 the cross section for the ALP-photon conversion
 in crystal is
 \bea
 \sigma &&=\frac{1}{2 E_i v_i} \int \frac{d^3 k_f}{(2\pi)^3}
 \frac{1}{2 E_f} 2 \pi \delta(E_i-E_f) \nnb \\
 && \times \bigg| \int_\Omega d^3x ~g_\phi~
 ({\vec k}_\gamma \times {\vec E})\cdot {\hat \epsilon} 
 ~e^{i({\vec k}_i-{\vec k}_f)\cdot {\vec x}}\bigg|^2, \label{X-section}
 \eea
 where $E_i$ and $E_f$ are the energies of the initial and final 
 particles respectively, $v_i$ the velocity of the initial particle
 relative to the crystal detector,
 ${\vec k}_i$ and ${\vec k}_f$ the initial
 and final momenta, ${\vec E}={\vec E}({\vec x})$ the electric 
 field in the crystal, ${\vec k}_\gamma$ the momentum of the photon,
 ${\vec \epsilon}$ the polarization vector of photon:
 $\epsilon^\mu=(0,{\vec \epsilon})$. $\Omega$ is 
 the volume of the crystal. One can clearly see 
 in (\ref{X-section}) that photon polarized normal to
 the plane of ${\vec k}$ and ${\vec E}$ involve into the transformation
 with ALP. Photon polarized in the plane of ${\vec k}$ and ${\vec E}$
 does not interact with ALP. 

\begin{figure}[tb]
\begin{center}
\begin{tabular}{cc}
\includegraphics[scale=1,width=8cm]{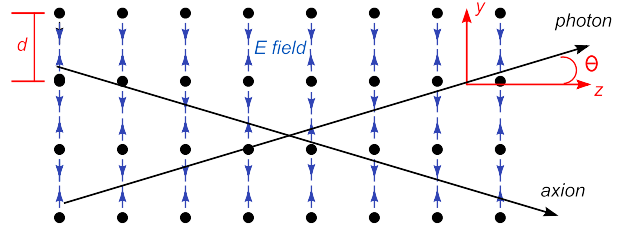}
\end{tabular}
\end{center}
\caption{ALP-photon conversion in intense crystalline field.
 Plane continuum electric field in y direction is plotted. } 
 \label{photon-crystal}
\end{figure}

 If the path length in crystal is a constant the cross-section can be expressed as
 $\sigma=S P$ where $P$ is the probability of ALP-photon conversion
 and $S$ is the geometric cross section of the target crystal.  
 A complete expression of $P$ should sum contributions of all
 ${\vec E}_{\vec q}$ in crystal.
 For simplicity we concentrate on the case that ALP interact efficiently
 with a plane electric field which is periodic in $y$ direction
 and constant in $x$ and $z$ direction, as shown in Fig.\ref{photon-crystal}.
 The relevant reciprocal vector is perpendicular to the
 crystallographic plane:
 ${\vec q}_T=q_T {\hat y}$ with $q_T=2 \pi n_T/d$ where
 $n_T\neq 0$ is an integer.

\begin{figure}[tb]
\begin{center}
\begin{tabular}{cc}
\includegraphics[scale=1,width=5.5cm]{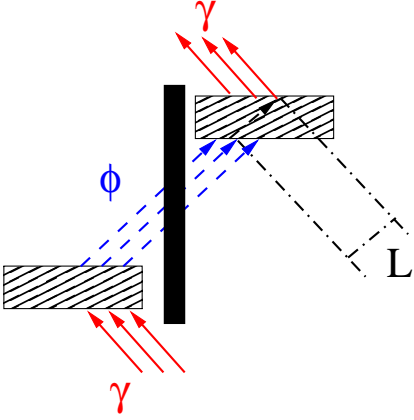}
\end{tabular}
\end{center}
\caption{Reflecting through the wall.
Photons are sent to crystal. Axion-like particles produced
in crystal are reflected by the crystal and after passing
through the wall generate photons in another crystal target.
} \label{through-wall}
\end{figure}

 Implementing (\ref{e-field}) into (\ref{X-section})
 and integrating over x and y coordinate
 we find that coherent ALP-photon conversion happens for 
  \bea
 k_f^x=k_i^x=0, ~k_i^y-k_f^y-q_T=0.
\label{cond2}
 \eea
Then after integration over z coordinate we find
 \bea
 P=(\frac{1}{2} g_\phi |E_T| L \cos\theta_\gamma \cos\theta_i)^2 G
 \frac{\sin^2(\Delta L/(4 |{\vec k}_i|))}{(\Delta L/(4 |{\vec k}_i|))^2},
 \label{prob}
 \eea
 where $L$ is the path length in crystal as shown in
 Fig. \ref{through-wall}.
 $\theta_\gamma$ is the angle between photon direction and
 z axis as shown in Fig. \ref{photon-crystal}. $\theta_i$ is the
 angle of the initial particle. For $\gamma \to \phi$ process,
 $\theta_i=\theta_\gamma$.
 $\Delta$ is
 \bea
 \Delta=m^2_f -m^2_i-2 q_T(k^y_i-\frac{1}{2} q_T),
 \label{Delta}
 \eea
 where $m_f$ and $m_i$ are masses of the final and initial particles.
 For $\gamma \to \phi$ process, $m_f=m_\phi$ and $m_i=\omega_p$ where 
 $\omega_p$ is the plasma frequency of photon in crystal
 and is around tens eV. 
 $G=1-\omega_p^2/E^2$ where $E$ is energy. 
 $G$ can be taken as one for $\omega_p \ll E$.
 $E_T= -i q_T U_{q_T}$ is the Fourier transform of the electric field 
 in $y$ direction. Using (\ref{potential2}) $E_T$ is expressed as
 \bea
 E_T= -i q_T \frac{1}{d} \int_0^d ~dy ~U(y) e^{i q_T y} \label{e-field2}.
 \eea
 For example, for $(110)$ plane of W crystal and $q_T=2\pi/d$
 we find $|E_T|=1.7 \times 10^{11}$ V$/$cm.
 For $(110)$ plane of the Ge crystal and $q_T=2\pi/d$ 
 we find $|E_T|=4.0 \times 10^{10}$ V$/$cm. 
 We see that when $|\Delta L/(4|{\vec k}_i|)| >1$
 the conversion probability is oscillatory 
 with an amplitude $4 g_\phi^2 |E_T|^2 |{\vec k}_i|^2 \cos^2\theta_\gamma
 \cos^2\theta_i /\Delta^2$.
 When $|\Delta L/(4|{\vec k}_i|)| \lsim 1$ the probability is resonantly
 enhanced with a value
 about $\frac{1}{4} g^2_\phi |E_T|^2 L^2 \cos^2\theta_\gamma \cos^2\theta_i$. 

 It's easy to find that $\Delta =0$ 
 corresponds to $k^z_f=k^z_i$ and together with (\ref{cond2})
 we see that the resonance limit corresponds to the case with
 \bea
 {\vec k}_i-{\vec k}_f-{\vec q}_T = 0 \label{cond1}.
 \eea
 In this limit the component ${\vec E}_{{\vec q}_T}$ associated with
 ${\vec q}_T$ in (\ref{e-field}) add coherently and give
 an resonant enhancement to the cross-section.
 When this resonant enhancement happens the cross-section
 is proportional to $|{\vec E}_{{\vec q}_T}|^2$ and 
 only photon polarized in ${\vec k}_\gamma \times {\vec q}_T$
 direction involves into the coherent conversion with ALP.
 In the resonance region ${\vec E}_{\vec q}$ associated with 
 all other reciprocal vectors
 effectively contribute zero after integration over space.

 In the resonant limit we find that
 \bea
 && k^y_i=\frac{1}{2} q_T+\frac{m^2_f-m_i^2}{2 q_T},
  \label{cond3} \\
 && k^y_f=-\frac{1}{2} q_T+\frac{m_f^2-m_i^2}{2 q_T}.
 \label{cond3b}
 \eea
 In this limit the incident angle and outgoing angle are
 \bea
 \sin\theta_i^0= \frac{q_T}{2|{\vec k}_i|}
 +\frac{m_f^2-m_i^2}{2 |{\vec k}_i| q_T},
 \label{angle1}\\
 \sin\theta_f^0= -\frac{q_T}{2 |{\vec k}_f|}
 +\frac{m_f^2-m_i^2}{2 |{\vec k}_f| q_T} .
 \label{angle1b}
 \eea
 We see that due to the periodic electric field in crystal 
 the final particle is reflected by the crystal.
 The resonant conversion requires $|\Delta L/(4E)|\lsim 1$ and we find 
 accordingly that the width of resonance in energy and incident angle is:
 $|\delta E| \lsim 2 |{\vec k}_i| L^{-1}/(q_T \sin\theta_i^0)$
 and $|\delta \theta_i| \lsim 2 E L^{-1}/(q_T |{\vec k}_i| \cos\theta_i^0)$.
 
 Note that in the limit $(\omega_p, m_\phi) \ll (q_T, E)$ or
 if $m_\phi \approx \omega_p$,
 (\ref{angle1}) and (\ref{angle1b}) are reduced to
 the Bragg scattering angle:
 \bea
 && \sin\theta_i^0 =\frac{ q_T}{2 |{\vec k}_i|},
 ~~\sin\theta_f^0=-\frac{q_T}{2 |{\vec k}_f|}.
 \label{angle2} 
 \eea
 In general case (\ref{angle1}) and (\ref{angle1b}) are required
 to be away from the Bragg scattering angle.
 
 We stress that according to (\ref{angle1}) and (\ref{angle1b})
 a careful arrangement of the incident angle
 with energy is required for resonant ALP-photon transformation 
 to happen.
 Another condition is that the energy $E$ should be larger than
 $q_T/2$:
 \bea
 E > \frac{\pi}{d}= 3.09 ~\textrm{keV} \times ~\frac{0.2 ~\textrm{nm}}{d}.
 \label{condE}
 \eea
 
 Solar axions have average energy $\langle E \rangle \approx 4.2$ keV
 ~\cite{solar} and have a significant fraction of flux with $E \gsim 3$ keV . 
 So crystal detector, e.g. making use of $(110)$ plane in Ge 
 crystal described above, can be used to detect solar axion.
 The experiment can be done by carefully arranging the crystal target
 such that the solar axion
 flux incident on the crystal with a specific angle to a chosen
 crystallographic plane.
 Due to the condition (\ref{angle1}),
 for a chosen angle only axions of selected energy contribute to the
 resonant ALP-photon conversion. 
 The spectrum of solar axion can be scanned by adjusting the angle of 
 crystal target to solar axion flux. Note that for a different crystallographic
 plane the resonance condition may also be satisfied
 for a different energy. Since the spectrum
 of solar axion is continuous many crystallographic planes can
 contribute to the resonant ALP-photon conversion.
 Detailed proposal of solar axion search experiments using
 crystalline field should take all possible planes
 into account. This is beyond the scope of this
 Letter
 
 Another interesting experiment is a reflecting-through-wall
 experiment, a variety of the 
 shining-through-wall experiment~\cite{thr-wall}. The experimental setup 
 can be arranged as shown in Fig. \ref{through-wall}. Photons such as
 hard X-rays which have large enough penetration depth in crystal
 are sent to a crystal. Due to the condition of the
 resonant conversion ALPs produced are reflected by the crystal.
 After passing through a wall ALPs convert back to photons in crystalline field
 of another crystal target. This experiment can be done using 
 mono-energetic X-rays with careful angular arrangement of crystal
 targets and incident angle of X-rays.
 In a symmetric experimental setup the probability finding X-rays 
 through the wall is 
 \bea
 P=(\frac{1}{2} g_\phi |E_T| L \cos^2\theta_\gamma )^4
 \frac{\sin^4(\Delta L/(4 E))}{(\Delta L/(4 E))^4}.
 \label{prob2}
 \eea
 (\ref{prob2}) is estimated as
 \bea
 &&P= 7.5 \times 10^{-20} \times \cos^4\theta_\gamma 
 \frac{\sin^4(\Delta L/(4 E))}{(\Delta L/(4 E))^4} \nnb \\
 &&\times \bigg( \frac{|E_T|}{2. \times 10^{11} \frac{\textrm{V}}{\textrm{cm}}} \bigg)^4
 \bigg( \frac{g_\phi}{10^{-8} \textrm{GeV}^{-1}} \bigg)^4
 \bigg( \frac{l}{5 \textrm{cm}} \bigg)^4,~~~~~~
 \label{prob3}
 \eea
 where $l=L \cos\theta_\gamma$ is the length of crystal.
 The event rate is
 \bea
 N= 19.7 ~\textrm{year}^{-1} \frac{{\cal W}}{10 ~\textrm{W}} 
 \frac{10 ~\textrm{keV}}{E} \frac{F}{1 \%}
 \frac{P}{10^{-20}}, \label{event}
 \eea
 where ${\cal W}$ is the power of the X-ray source, 
 $F$ the efficiency
 of the coherent conversion of ALP and X-rays.
 $F$ arises due to
 the fact that X-ray source has an angular spread and energy spread
 and only parts of X-rays can satisfy the condition of resonant conversion:
 $|\Delta L/(4E)| \lsim 1$.
 We can see that this experiment is very interesting.
 If $1\%$ efficiency can be achieved with a X-ray source of
 $10$ W  power,
 a detector using the $(110)$ plane of W crystal with $5$cm length
 can probe $g_\phi$ to $10^{-8}$ GeV$^{-1}$. It would be much better than
 the shining-through-wall experiment using B field~\cite{review1}.
 
 A very interesting aspect of ALP search experiment using crystal
 is that the range of $m_\phi$ can be scanned by varying the incident
 angle. For example, 
 for a fixed incident angle $\theta_i$ in $\phi \to \gamma$ process 
 the resonance($\Delta=0$) happens for $m_\phi$:
 \bea
 m_\phi^2=\omega_p^2-2 q_T( |{\vec k}_i|\sin\theta_i- \frac{1}{2} q_T) \label{scan}.
 \eea
 $m_\phi$ range can be
 scanned in experiment by carefully adjusting the incident angle
 $\theta_i$. This is an
 important virtue of the ALP search experiment using crystal. Using
 hard X-rays or the solar axion source, ALP search experiments
 using crystalline field can probe the range of $m_\phi$ up to 
 keV scale without loss of sensitivity.

 In conclusion we find that coherent ALP-photon conversion
 can happen in crystalline field if the energy is larger than
 about keV and a condition of the energy and the incident angle
 is satisfied. We propose to do
 ALP search experiments using intense electric field in crystal.
 These experiments have the following virtues: 1) 
 Since the crystalline field is several orders of magnitude
 stronger than the magnetic field available in laboratory, 
 ALP search experiment using crystalline field has the potential
 to reach sensitivity beyond the present experiments based on
 ALP-photon conversion in magnetic field. 2)
 ALP search experiments using crystalline field can probe wide
 range of $m_\phi$, from eV to keV,
 by adjusting the incident angle of
 initial flux to crystal
 when using hard X-ray or using solar axion source in experiments. 
 This range of $m_\phi$ cannot be probed with good sensitivity in
 ALP search experiments using external B field. 

 {\bf Acknowledgment:} I would like to thank X. Y. Li, Y. Liao and
 Y. Q. Chen for fruitful discussions on physics in intense external fields.
 This work is supported by Science and Technology Commission of Shanghai 
 Municipality under contract No. 09PJ1403800 and
 National Science Foundation of China(NSFC), grant No. 10975052.

 \vskip 0.1cm
 {\it Note added} ~After the submission of this Letter we received Refs. 
 ~\cite{early1,early2,exp1,exp2,exp3,exp4,exp5}. The idea of photon-axion 
 conversion in crystal has been studied in ~\cite{early1,early2} and 
 solar axion search experiments based on ~\cite{early2} have been done
 ~\cite{exp1,exp2,exp3,exp4,exp5}. 
 We note that the result
 in this Letter is different from the result in \cite{early1,early2}:
 1) The probability (\ref{prob}) has a different dependence
 on the angle $\theta_\gamma$ than in ~\cite{early1,early2}.
 2)
 The result in this Letter is valid for all possible angles but the result
 in ~\cite{early1,early2} works only for the Bragg scattering angle which is a 
 special case of our result, as shown in (\ref{angle2}).
 3) The resonance condition for non-zero mass case, (\ref{angle1}),
 was not found in ~\cite{early1,early2}. According to result in this Letter
 previous experiments~\cite{exp1,exp2,exp3,exp4,exp5} based on ~\cite{early2}
 should be re-examined.

\end{document}